\documentclass[preprint,floats,tightenlines,11pt,eqsecnum,aps]{revtex4}
\usepackage{amsmath}
\usepackage{graphicx}
\begin{document}
\def\appls{\hbox{$<$\kern-.75em\lower 1.00ex\hbox{$\sim$}}}
\title{VIOLATION OF GENERALIZED BOSE-EINSTEIN SYMMETRY AND\\
QUANTUM ENTANGLEMENT OF $\pi^- \pi^+$ ISOSPIN STATES\\ IN PION PAIR PRODUCTION $\pi N \to \pi^- \pi^+ N$}

\author{Miloslav Svec\footnote{electronic address: svec@hep.physics.mcgill.ca}}
\affiliation{Physics Department, Dawson College, Montreal, Quebec, Canada H3Z 1A4}
\date{October 16, 2007}

\begin{abstract}
Generalized Bose-Einstein symmetry requires that $J+I = even$ for two-pion angular states of spin $J$ and total isospin $I$. We show that the symmetry predicts three linearly independent constraints on partial wave intensities with even spin for $\pi^- p \to \pi^- \pi^+ n$, $\pi^- p \to \pi^0 \pi^0 n$ and $\pi^+ p \to \pi^+ \pi^+ n$. Available data violate all three constraints for $S$-, $D^0$-, $D^U$- and $D^N$-partial waves. The violations of the symmetry imply a presence of the symmetry violating contributions in transversity amplitudes in $\pi^- p \to \pi^- \pi^+ n$ and predict quantum entanglement of $\pi^- \pi^+$ isospin states which is excluded by the symmetry. We derive approximate lower and upper bounds on entanglement amplitudes $|a_S|$ and $|a_A|$. The bounds provide a clear evidence for entanglement of $\pi^- \pi^+$ isospin states below 840 MeV and suggest the entanglement at higher dipion masses. The small values of $|a_S| \sim 0.10-0.20$ below 840 MeV explain the puzzling differences between the $S$-wave intensities in $\pi^- p \to \pi^- \pi^+ n$ and $\pi^- p \to \pi^0 \pi^0 n$ and reveal a suppression of isospin $I=0,2$ contribution in the $S$-wave amplitudes in $\pi^- p \to \pi^- \pi^+ n$. The large isospin $I=1$ contribution of $\rho^0(770)$ to both $S$- and $P$-wave amplitudes is due to large entanglement amplitude $|a_A| \sim 0.98-0.99$. These findings confirm the predictions of a model of non-unitary dynamics of the pion creation processes arising from a $CPT$ violating interaction of these processes with a quantum environment.

\end{abstract}

\maketitle

\tableofcontents

\newpage

\section{Introduction.}

Measurements of pion creation processes 
$\pi^- p \to \pi^- \pi^+ n$~\cite{becker79a,becker79b,chabaud83,rybicki85,alekseev99} and $\pi^+ n \to \pi^+ \pi^- p$~\cite{lesquen85} on polarized targets enable model independent determination of production amplitudes. Evidence for a narrow rho-like resonance in the $S$-wave transversity amplitudes was found in amplitude analyses of CERN data on $\pi^- p \to \pi^- \pi^+ n$ at 17.2 GeV/c at small momentum transfers~\cite{becker79b,chabaud83,donohue79,svec92b,svec96,svec97a,svec02a} as well as at large $t$~\cite{rybicki85}, in analysis of ITEP data on the same reaction at 1.78 GeV/c at small $t$~\cite{alekseev99}, and in the analyses of CERN data on $\pi^+ n \to \pi^+ \pi^- p$ at 5.98 and 11.85 GeV/c at larger momentum transfers~\cite{svec84,svec92b,svec96}. The $S$- and $P$-wave subsystem of the reduced density matrix measured on transversely polarized target is analytically solvable at any dipion mass~\cite{svec07b,svec07c}. The moduli of the $S$-wave transversity amplitudes $|S_\tau|^2$ are given by a simple expression
\begin{equation}
|S_\tau|^2=a_{1,\tau}+a_{2,\tau}-3|L_\tau|^2
\end{equation}
where $\tau=u,d$ are target nucleon transverse spins "up" and "down" relative to the scattering plane. The terms $a_{1,\tau}+a_{2,\tau}$ are expressed in terms of measured spin density matrix elements and exhibit a clear $\rho^0(770)$ resonant peak~\cite{svec07b} that survives the subtraction of the $\rho^0(770)$ peak in the $P$-wave amplitude $|L_\tau|^2$. The rho-like resonance present in the $S$-wave amplitudes  - previously referred to as $\sigma(750)$ resonance - must be interpreted as
$\rho^0(770)$ resonance. We conclude that the data on polarized targets reveal $\rho^0(770)-f_0(980)$ mixing in the $S$-wave amplitudes which appears to violate Lorentz symmetry and Generalized Bose-Einstein symmetry for pion isospin multiplet.\\

The origin of $\rho^0(770)-f_0(980)$ mixing is related to another facet of the CERN data on polarized target. In the previous work~\cite{svec07a} we showed that the data at large momentum transfers $t$ provide evidence for evolution of pure initial states to mixed final states in $\pi^- p \to \pi^- \pi^+ n$. In quantum theory such non-unitary evolution occurs in open quantum systems $S$ interacting with a quantum environment $E$. While the system $S$ and the environment $E$ undergo a unitary co-evolution $\rho_f(S,E)=U\rho_i(S,E)U^+$, the evolution of the reduced density matrix 
$\rho_f(S)=Tr_{E}\rho_f(S,E)=\mathcal{E}(\rho_i(S))$ of the system $S$ is non-unitary and is given by Kraus representation~\cite{kraus83,nielsen00,breuer02,bengtsson06}
\begin{equation}
\rho_f(S)=\mathcal{E}(\rho_i(S))=\sum \limits_\ell \sum \limits_{m,n} p_{mn} S_{\ell m} \rho_i(S)S^+_{n \ell}
\end{equation}
where $S_{\ell m}=<e_\ell|U|e_m>$ and $|e_\ell>$ are interacting degrees of freedom of the environment. The initial state of the environment has a general form
\begin{equation}
\rho_i(E)=\sum \limits_{m,n} p_{mn}|e_m><e_n|
\end{equation}
We showed in~\cite{svec07a} that Kraus representation leaves invariant the formalism used in data analyses provided the co-evolution with the environment conserves $P$-parity and quantum numbers of the environment. On general grounds~\cite{nielsen00,svec07a} there are four interacting degrees of freedom of the environment in $\pi^- p \to \pi^- \pi^+ n$. The Kraus representation then has a diagonal form
\begin{equation}
\rho_f(S)=\sum \limits_{\ell=1}^4 p_{\ell \ell} S_{\ell \ell} \rho_i(S)S^+_{\ell \ell}= \sum \limits_{\ell=1}^4 p_\ell \rho_f(\ell)
\end{equation}

The question now arises what are the quantum states of the environment in $\pi^- p \to \pi^- \pi^+n$. In~\cite{svec07b} we associate with the two solutions for transversity amplitudes $A_u(i)$ and $A_d(j)$ two one-qubit states $|i>$ and $|j>$ where $i,j=1,2$ and identify the quantum states $|e_\ell>$ with four two-qubit states $|i>|j>$. The final state $\rho_f(S)$ is mixed state of four solutions $\rho_f(ij)$ with probabilities $p_\ell \equiv p_{ij}$ and $\sum \limits_{i,j=1,2} p_{ij}=1$. The measured final state $\rho_f(\pi^-\pi^+n)$ is mixed state even when the initial states $\rho_i(\pi^-p)$ are pure states. According to Wald Theorem~\cite{wald80}, the interaction giving rise to such a non-unitary evolution from pure states to mixed states must violate $CPT$ symmetry. The violation of the $CPT$ symmetry implies that such interaction violates also Lorentz symmetry~\cite{greenberg02}.\\

In our previous work~\cite{svec07b} we have formulated a model of non-unitary dynamics of $\rho^0(770)-f_0(980)$ mixing. According to the model, the pion creation process $\pi^- p \to \pi^- \pi^+ n$ proceeds in three stages. In the first stage resonant and non-resonant (isospin 2) $q \overline{q}$ ($q\overline{q}q \overline{q}$) modes with spin $K$, helicity $\mu$ and isospin $I_K$ are formed in a process $\pi^- p \to q \overline{q} n$ that preserves Lorentz and $CPT$ symmetry of QCD dynamics. In the second stage the resonant modes propagate with Breit-Wigner amplitudes. As the modes propagate they interact in the last stage with the environment states $|i>|j>$. This $CPT$ violating interaction has two effects. First, the transversity amplitudes $A_u$ and $A_d$ split into two solutions each $A_u(i), i=1,2$ and $A_d(j),j=1,2$ with the solution labels now re-interpreted as quantum numbers of the environment. Second, the interaction induces isospin conserving transitions between $q \overline{q}$ states of spin $K$, helicity $\mu$ and isospin $I_K$ into two-pion states with spin $J$, helicity $\lambda$ and isospin $I(\pi^- \pi^+)=I_K$.  As a result, $\rho^0(770)$ appears in the $S$-wave amplitudes and $f_0(980)$ appears in the $P$-wave amplitudes, as observed experimentally~\cite{svec07b}.\\

Generalized Bose-Einstein symmetry extends the Bose-Einstein statistics for identical bosons to the entire isospin meson multiplet~\cite{martin70}. As the result, two-meson states must have a definite symmetry under permutation of the members of the multiplet. In the case of two-pion states the Generalized Bose-Einstein symmetry requires that $J+I=even$ where $J$ and $I$ are spin and isospin of the two-pion state~\cite{martin70}.\\

The environment induced transitions between $q \overline{q}$ modes and two-pion states violate Generalized Bose-Einstein symmetry. Transversity amplitudes $A^{J,\eta}_{\lambda,\tau}$ are then a superposition of Generalized Bose-Einstein symmetry conserving and violating amplitudes $C^{J,\eta}_{\lambda,\tau}$ and $V^{J, \eta}_{\lambda, \tau}$
\begin{eqnarray}
A^{J,\eta}_{\lambda,u}(i) & = & \alpha_{J\lambda, \eta} C^{J,\eta}_{\lambda,u}(i)+\omega_{J\lambda, \eta} V^{J,\eta}_{\lambda,u}(i)\\
\nonumber
A^{J,\eta}_{\lambda,d}(j) & = & \alpha_{J\lambda ,\eta} C^{J,\eta}_{\lambda,d}(j)+\omega_{J\lambda, \eta} V^{J,\eta}_{\lambda,d}(j)
\end{eqnarray}
where $\eta$ is $t$-channel naturality of the amplitude~\cite{svec07a}. The self-consistency of (1.5) with angular expansion of the total amplitude requires~\cite{svec07b}
\begin{eqnarray}
\text{J=even:} & \alpha_{J\lambda,\eta}=a_S, & \omega_{J\lambda, \eta}=a_A\\
\nonumber
\text{J=odd:} & \alpha_{J\lambda,\eta}=a_A, & \omega_{J\lambda, \eta}=a_S
\end{eqnarray}
where $|a_S|^2+|a_A|^2=1$. The novel feature of the model is its prediction that the produced two-pion states are not separable but entangled isospin states
\begin{equation}
|E(\pi^-\pi^+)>=a_S|S>+a_A|A>=a|\pi^-\pi^+>+b|\pi^+\pi^->
\end{equation}
where $|S>$ and $|A>$ are symmetric and antisymmetric $\pi^-\pi^+$ isospin states, respectively, and
\begin{eqnarray}
a & = {1\over{\sqrt{2}}}(a_S+a_A)\\
\nonumber
b & = {1\over{\sqrt{2}}}(a_S-a_A)
\end{eqnarray}
Only for $a_S=a_A={1\over{\sqrt{2}}}$ are the two-pion states $\pi^-\pi^+$ separable.

The assumption of Generalized Bose-Einstein symmetry implies a relation between transversity amplitudes for even spin $J$ in $\pi^- p \to \pi^- \pi^+ n$, $\pi^- p \to \pi^0 \pi^0 n$ and $\pi^+ p \to \pi^+ \pi^+ n$
\begin{equation}
A^{J,\eta}_{\lambda,\tau}(-+)=A^{J,\eta}_{\lambda,\tau}(00)-
\sqrt{{3\over{2}}}A^{J,\eta}_{\lambda,\tau}(++) 
\end{equation}
The symmetry thus imposes constraints on the observables in these three reactions. The purpose of this work is to use the available data to test the validity of the Generalized Bose-Einstein symmetry. In addition, we examine the data for evidence of quantum entanglement of $\pi^-\pi^+$ isospin states.\\

The paper is organized as follows. In Section II. we introduce Generalized Bose-Einstein symmetry of two-pion states in reactions $\pi N \to \pi \pi N$. In Section III. we derive three constraints on partial wave intensities with even spin in $\pi^- p \to \pi^- \pi^+ n$, $\pi^- p \to \pi^0 \pi^0 n$ and $\pi^+ p \to \pi^+ \pi^+ n$ which are the consequence of Generalized Bose-Einstein symmetry. In Section IV. we gather data required for the tests of the constraints and present the results of the tests. The available data violate the symmetry in all partial waves tested. In Section V. we derive approximate lower and upper bounds on entanglement amplitudes $|a_S|$ and $|a_A|$. The bounds provide a clear evidence for quantum entanglement of $\pi^- \pi^+$ isospin states below 840 MeV and suggest the entanglement at higher dipion masses. The small values of $|a_S| \sim 0.10-0.20$ below 840 MeV explain the puzzling difference between the $S$-wave intensities in $\pi^- p \to \pi^- \pi^+ n$ and $\pi^- p \to \pi^0 \pi^0 n$ and reveal a suppression of isospin $I=0,2$ contribution in the $S$-wave amplitudes in $\pi^- p \to \pi^- \pi^+ n$. The large isospin $I=1$ contribution of $\rho^0(770)$ to both $S$- and $P$-wave amplitudes is due to large entanglement amplitude $|a_A| \sim 0.98-0.99$. The paper closes with a summary and concluding remarks in Section VI.

\section{Generalized Bose-Einstein symmetry of two-pion states in\\$\pi N \to \pi \pi N$ reactions.}

A state consisting of two identical particles must be described by a state vector with definite symmetry properties with respect to interchange of the particles. The two-particle state must be symmetric if the particles are bosons and antisymmetric if they are fermions. Consider two-body angular momentum state $|JM;\mu_1\mu_2>$ where $J$ is the total spin and $\mu_1$ and $\mu_2$ are helicities of the particles. The properly symmetrized or antisymmetrized physical state is given by~\cite{martin70}
\begin{equation}
N(|JM;\mu_1 \mu_2>+(-1)^J|JM;\mu_2 \mu_1>)
\end{equation}
where the normalization coefficient $N=\sqrt{1\over{2}}$ if $\mu_1 \neq \mu_2$ and $N={1\over{2}}$ if $\mu_1=\mu_2$. Bose-Einstein statistics requires that two-pion states $\pi^0 \pi^0$ and $\pi^+ \pi^+$ must have even $J$.\\

Generalized Bose-Einstein symmetry is an extension of Bose-Einstein statistics to all particles belonging to an isospin multiplet of isospin $I_1$ which are regarded as $2I_1+1$ charge states of the same particle. By incorporating the isospin quantum numbers in the state vector the symmetrization properties are extended to the interchange of particles belonging to the same multiplet. Then the properly symmetrized or antisymmetrized angular momentum state of total isospin $I$ of two particles belonging to the same multiplet $I_1$ is given by~\cite{martin70}
\begin{equation}
|JM;\mu_1\mu_2;Im>+(-1)^{J+I-2I_1}|JM;\mu_2\mu_1;Im>
\end{equation}
If $\mu_1=\mu_2$ then $J+I-2I_1$ must be even. Generalized Bose-Einstein symmetry requires that for two-pion state $J+I$ is even.\\

Consider separable two-pion state $|\pi^->|\pi^+>$. It can be written in the form
\begin{equation}
|\pi^-\pi^+>={1\over{\sqrt{2}}}|S>+{1\over{\sqrt{2}}}|A>
\end{equation}
where $|S>$ and $|A>$ are symmetric and antisymmetric maximally entangled $\pi^-\pi^+$ isospin states
\begin{eqnarray}
|S> & = & {1 \over{\sqrt{2}}}(|\pi^->| \pi^+>+|\pi^+>|\pi^->)=
-{1 \over {\sqrt{3}}}(\sqrt{2}|0,0>+|2,0>)\\
\nonumber
|A> & = & {1 \over{\sqrt{2}}}(|\pi^->| \pi^+>-|\pi^+>|\pi^->) = |1,0>
\end{eqnarray}
where we used the convention $|\pi^+>=-|1,+1>$~\cite{gibson76}. Transversity amplitudes are matrix elements
\begin{equation}
A^{J\eta}_{\lambda,\tau}=<J\lambda\eta|<\pi^-\pi^+|<\tau_n|T|0\tau>=
{1\over{\sqrt{2}}}C^{J\eta}_{\lambda,\tau}+
{1\over{\sqrt{2}}}V^{J\eta}_{\lambda,\tau}
\end{equation}
where $\tau$ and $\tau_n$ are target and recoil nucleon transversities and 0 stands for incident pion helicity. The transversity amplitudes are equal to Generalized Bose-Einstein symmetry conserving amplitudes $C^{J\eta}_{\lambda,\tau}$
\begin{eqnarray}
\text{J=even:} \quad C^{J,\eta}_{\lambda,\tau} & = & <J\lambda \eta|<S|<\tau_n|T|0\tau>\\
\nonumber
\text{J=odd:} \quad C^{J,\eta}_{\lambda,\tau} & = & <J\lambda \eta|<A|<\tau_n|T|0\tau>
\end{eqnarray}
since the Generalized Bose-Einstein symmetry violating amplitudes $V^{J\eta}_{\lambda,\tau}$ vanish
\begin{eqnarray}
\text{J=even:} \quad V^{J,\eta}_{\lambda,\tau} & = & <J\lambda\eta|<A|<\tau_n|T|0\tau>=0\\
\nonumber
\text{J=odd:} \quad V^{J,\eta}_{\lambda,\tau} & = & <J\lambda \eta|<S|<\tau_n|T|0\tau>=0
\end{eqnarray}
The two-pion states $|\pi^0 \pi^0>$ and $|\pi^+ \pi^+>$ satisfy the constraint $J+I$ = even since they correspond to even $I$
\begin{eqnarray}
|\pi^0 \pi^0> & = & -{1\over{\sqrt{3}}}(|0,0>-\sqrt{2}|2,0>)\\
\nonumber
|\pi^+\pi^+> & = & |2,2>
\end{eqnarray}

Generalized Bose-Einstein symmetry is assumed to be a symmetry of strong interactions. It is violated by electromagnetic interactions. The observation of $\rho^0(770)-f_0(980)$ mixing suggests it is violated also in the interactions of the pion creation process with the quantum environment.

\section{Constraints on partial wave intensities from Generalized\\ Bose-Einstein symmetry.}

We first consider $S$-wave amplitudes $S_\tau(-+)$, $S_\tau(00)$ and $S_\tau(++)$ in  three measured processes $\pi^- p \to \pi^- \pi^+ n$, $\pi^- p \to \pi^0 \pi^0 n$ and $\pi^+ p \to \pi^+ \pi^+ n$
\begin{eqnarray}
S_\tau(-+) & = & <J=\lambda=0,\eta=-1|<\pi^-\pi^+|<\tau_n|T|0\tau>\\
\nonumber
S_\tau(00) & = & <J=\lambda=0,\eta=-1|<\pi^0\pi^0|<\tau_n|T|0\tau>\\
\nonumber
S_\tau(++) & = & <J=\lambda=0,\eta=-1|<\pi^+\pi^+|<\tau_n|T|0\tau>
\end{eqnarray}
Generalized Bose-Einstein symmetry requires
\begin{equation}
S_\tau(-+)={1 \over{\sqrt{2}}}<J\lambda \eta|<S|<\tau_n|T|0\tau>
\end{equation}
Using (2.4) and (2.8) we can express the amplitudes $S_\tau(c)$, $c=(-+),(00),(++)$ in terms of amplitudes $S_\tau^{II_3}$ with definite total isospin $I$ and $I_3$

\begin{subequations}
 \label{allequations} 
 \begin{eqnarray}
  S_\tau(-+)&=&-{1 \over{\sqrt{3}}} \{ S^{00}_\tau + {1 \over{2}}
  \sqrt{2}    S^{20}_\tau \} \label{equationa}
  \\
  S_\tau(00)&=&-{1 \over{\sqrt{3}}} \{ S^{00}_\tau - \sqrt{2}
  S^{20}_\tau\} \label{equationb}
  \\
  S_\tau(++)&=&S^{22}_\tau \label{equationc}
 \end{eqnarray}
\end{subequations}
Assuming that the $S$-matrix is invariant under the rotations in isospin space, the isospin amplitudes $S^{II_3}_\tau$ then do not depend on the  component $I_3$ and we have
\begin{equation}
S^{20}_\tau = S^{22}_\tau = S_\tau(++)
\end{equation}
From (3.3a) and (3.3b) we then get the equation (1.9)
\begin{equation}
S_\tau(00)=S_\tau(-+)+\sqrt{{3 \over{2}}} S_\tau(++)
\end{equation}
It is useful to write the following combinations of this equation
\begin{subequations}
 \label{allequations} 
 \begin{eqnarray}
  \sqrt{{3 \over{2}}} S_\tau(++)&=&S_\tau(00)-S_\tau(-+)
  \\
  S_\tau(-+)&=&S_\tau(00) - \sqrt{{3 \over{2}}} S_\tau(++)
  \\
  S_\tau(00)&=&S_\tau(-+) + \sqrt{{3 \over{2}}} S_\tau(++)
 \end{eqnarray}
\end{subequations}
and calculate the $S$-wave intensities
\begin{equation}
I_S(c)=|S_u(c)|^2 + |S_d(c)|^2 
\end{equation}
for $c=(++),(-+),(00)$ using expressions on r.h.s. of (3.6). With
\begin{equation}
I_S(2)={3 \over{2}}I_S(++)
\end{equation}
we then obtain
\begin{subequations}
 \label{allequations} 
 \begin{eqnarray}
  I_S(-+)+I_S(00) - I_S(2)&=&+2 \sum\limits_\tau
  Re[S_\tau(-+)S^*_\tau(00)] \\
  I_S(-+)-I_S(00) - I_S(2)&=&-2 \sqrt{{3 \over {2}}} \sum\limits_\tau
  Re[S_\tau(00)S^*_\tau(++)] \\
  I_S(-+)-I_S(00) + I_S(2)&=&-2 \sqrt{{3 \over {2}}} \sum\limits_\tau
  Re[S_\tau(-+)S^*_\tau(++)]
 \end{eqnarray}
\end{subequations}
The interference terms on the r.h.s. of (3.9) are scalar products of four-vectors
\begin{equation}
A_S(c)=\{Re S_u(c), Im S_u(c), Re S_d(c), Im S_d(c) \}
\end{equation}
in a Euclidian 4-dimensional space with the norm $A_S(c) \star A_S(c)=I_S(c)$ and scalar product
\begin{equation}
A_S(c) \star A_S(c') = \sqrt{I_S(c)} \sqrt{I_S(c')} \cos \Omega_{cc'}(S)
\end{equation}
where $\Omega_{cc'}(S)$ is an angle between the vectors $A_S(c)$ and $A_S(c')$. The relations (3.9) then read
\begin{subequations}
 \label{allequations} 
 \begin{eqnarray}
  I_S(-+)+I_S(00) - I_S(2)&=&+2 \sqrt{I_S(-+)} \sqrt{I_S(00)} \cos
  \Omega_1(S)
  \\
  I_S(-+)-I_S(00) - I_S(2)&=&-2 \sqrt{I_S(00)} \sqrt{I_S(2)} \cos
  \Omega_2(S)
  \\
  I_S(-+)-I_S(00) + I_S(2)&=&-2 \sqrt{I_S(-+)} \sqrt{I_S(2)} \cos
  \Omega_3(S)
 \end{eqnarray}
\end{subequations}
The equations (3.12) represent three linearly independent constraints on the measured spectra $I_S(-+)$, $I_S(00)$ and $I_S(++)= {2 \over{3}}I_S(2)$ imposed by the requirement that the cosines have physical values. While the cosines are linearly independent, they satisfy a non-linear constraint
\begin{equation}
\cos^2 \Omega_1(S) + \cos^2 \Omega_2(S) + \cos^2 \Omega_3(S) -2 \cos \Omega_1(S) \cos \Omega_2(S) \cos \Omega_3(S) = 1
\end{equation}
The constraint (3.13) implies that for physical values of the cosines the phases satisfy a phase condition
\begin{equation}
\Omega_1(S)+\Omega_2(S)+\Omega_3(S)=0
\end{equation}

There are 5 $D$-wave amplitudes in each process $c=(-+),(00),(++)$. The dominant $D^0_\tau(c)$ unnatural exchange amplitude corresponds to helicity $\lambda=0$. Then there are 2 unnatural exchange amplitudes $D^U_\tau(c)$ and $D^{2U}_\tau(c)$ and 2 natural exchange amplitudes $D^N_\tau(c)$ and $D^{2N}_\tau(c)$ corresponding to different combinations of helicities $\lambda=\pm1$ and $\pm2$, respectively. Relations arising from the assumption of Generalized Bose-Einstein symmetry among $D$-wave amplitudes for processes $c=(-+), (00),(++)$ are identical to relations (3.5) among the $S$-wave amplitudes. They give rise to constraints for $D$-wave intensities $I_{D^0}$, $I_{D^U}$, $I_{D^N}$ identical in form to the contstraints (3.12) for $S$-wave intensities. Measurements of pion pair production processes indicate that the helicity $\lambda = \pm 2$ amplitudes $D^{2U}_\tau$ and $D^{2N}_\tau$ are small and can be neglected at small momentum transfers $t$. While constraints analogous to (3.12) still hold, they will not be considered.\\

\section{Evidence for violation of Generalized Bose-Einstein symmetry.}

\subsection{Data used in the tests of Generalized Bose-Einstein symmetry.}

To test the constraints (3.12) we need data on the intensities $I_A(c)$ from $\pi^- \pi^+$, $\pi^0 \pi^0$ and $\pi^+ \pi^+$ production. The available data allow to perform the tests at small momentum transfers $0.005 < |t| < 0.20$ (GeV/c)$^2$.\\

For the $\pi^- \pi^+$ channel we used two analyses of the CERN measurements on polarized target at 17.2 GeV/c. One analysis is our high resolution analysis using Monte Carlo method presented in Ref.~\cite{svec07b}. It covers mass range 580 - 1080 MeV and involves no $D$-wave. Another analysis is the CERN-Cracow-Munich (CCM) analysis~\cite{kaminski02} which covers a larger mass region of 580-1620 MeV and includes $D$-wave above 980 MeV. Both analyses are similar below 1080 MeV and produce two solutions for the moduli $|S_u|^2$ and $|S_d|^2$. Two solutions for the $S$-wave intensity $I_S(-+)$ were used in this mass range corresponding to combinations (1,1) and (2,2) of solutions for the moduli
\begin{eqnarray}
I_S(-+) \text{ Solution (1,1)} & = & |S_u(1)|^2 +|S_d(1)|^2\\
\nonumber
I_S(-+) \text{ Solution (2,2)} & = & |S_u(2)|^2 +|S_d(2)|^2
\end{eqnarray}
There is only one solution for the intensities $I_A(-+)$, $A=S,D^0,D^U,D^N$ above 980 MeV in the CCM analysis. The results for $I_S(-+)$ from our analysis and the CCM analysis are shown in Figures 1 and 2, respectively. The results for the $D^0$-wave intensity from the CCM analysis are shown also in Figures 2. Figure 3 shows their results for $D^U$- and $D^N$-wave intensities. The intensities in the CCM analysis are in units $\mu b/20$ MeV. They were converted to units from Ref.~\cite{grayer74} used in our analysis using a conversion factor $0.109 \mu b/20$ MeV = 1000 events/20 MeV.\\

\begin{figure}
\includegraphics[width=14cm,height=12cm]{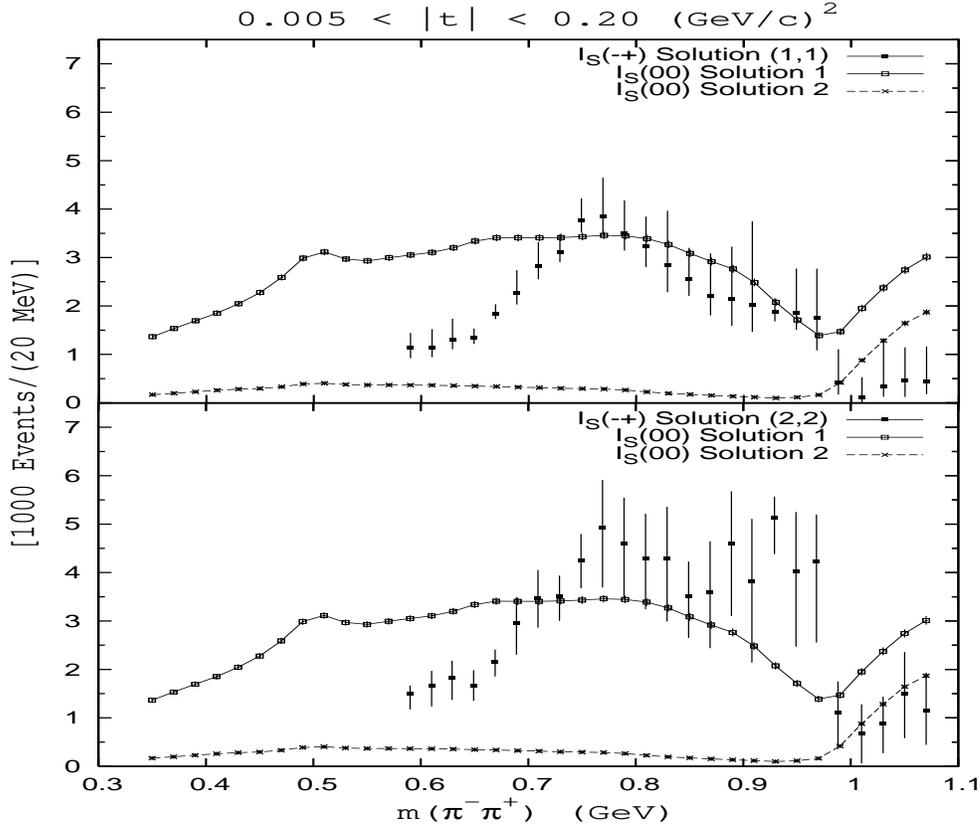}
\caption{Solutions (1,1) and (2,2) for $S$-wave intensity $I_S(-+)$ for $\pi^- p \to \pi^- \pi^+n$ from high resolution analysis~\cite{svec07b} compared with Solutions 1 and 2 for intensity $I_S(00)$ for $\pi^- p \to \pi^0 \pi^0 n$ from Ref.~\cite{gunter01}. The intensities $I_S(00)$ were interpolated to 20 MeV bins and scaled to 17.2 GeV/c.}
\label{fig1}
\end{figure}

\begin{figure}
\includegraphics[width=12cm,height=12cm]{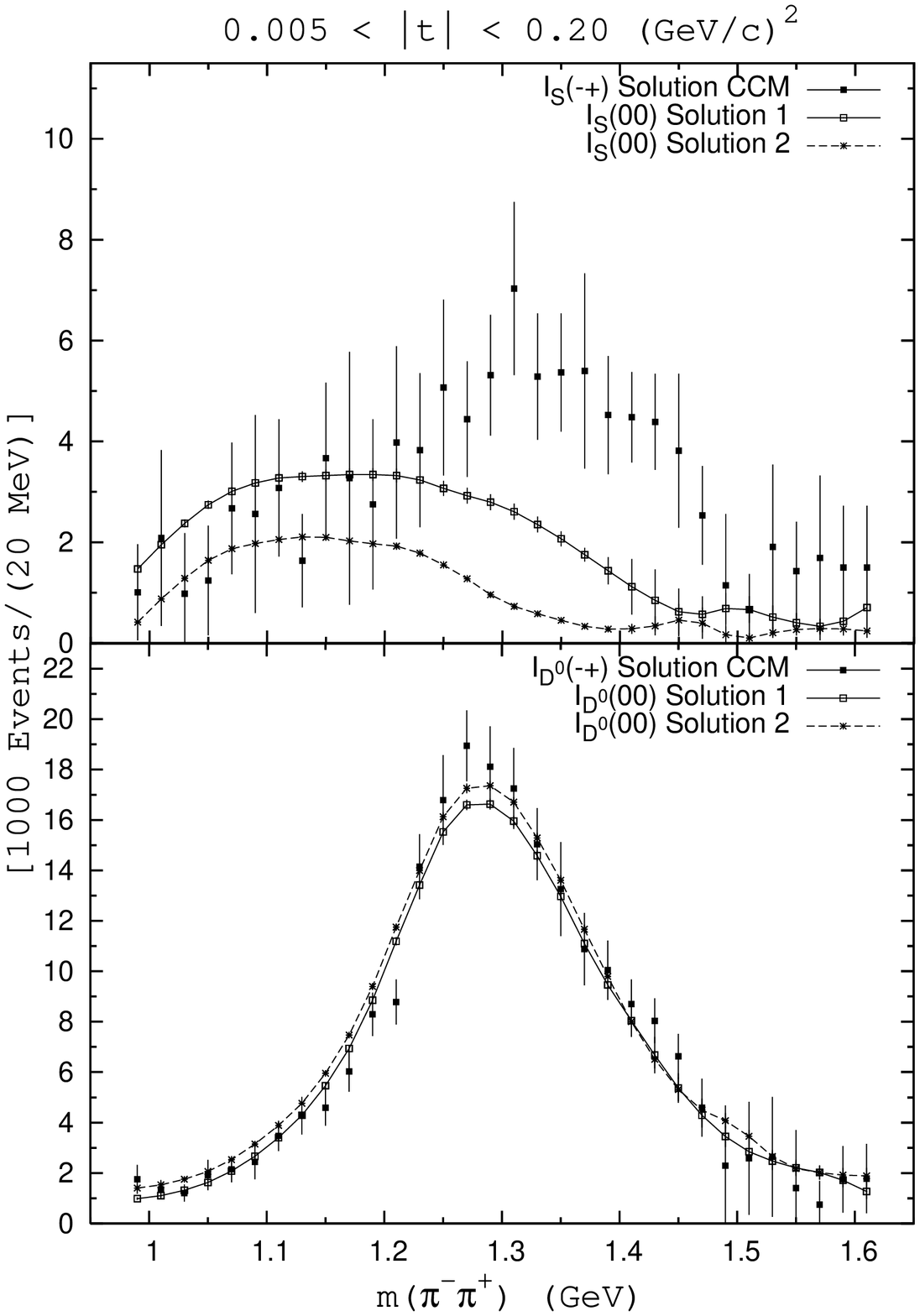}
\caption{$S$-wave and $D^0$-wave intensities $I_S(-+)$ and $I_{D^0}(-+)$ for $\pi^- p \to \pi^- \pi^+ n$ from CERN-Cracow-Munich analysis~\cite{kaminski02} compared to two Solutions 1 and 2 for intensities $I_S(00)$ and $I_{D^0}(00)$ for $\pi^- p \to \pi^0 \pi^0 n$ from Ref.~\cite{gunter01}. The intensities $I_S(00)$ and $I_{D^0}(00)$ were interpolated to 20 MeV bins and scaled to 17.2 GeV/c.}
\label{fig2}
\end{figure}

\begin{figure}
\includegraphics[width=12cm,height=12cm]{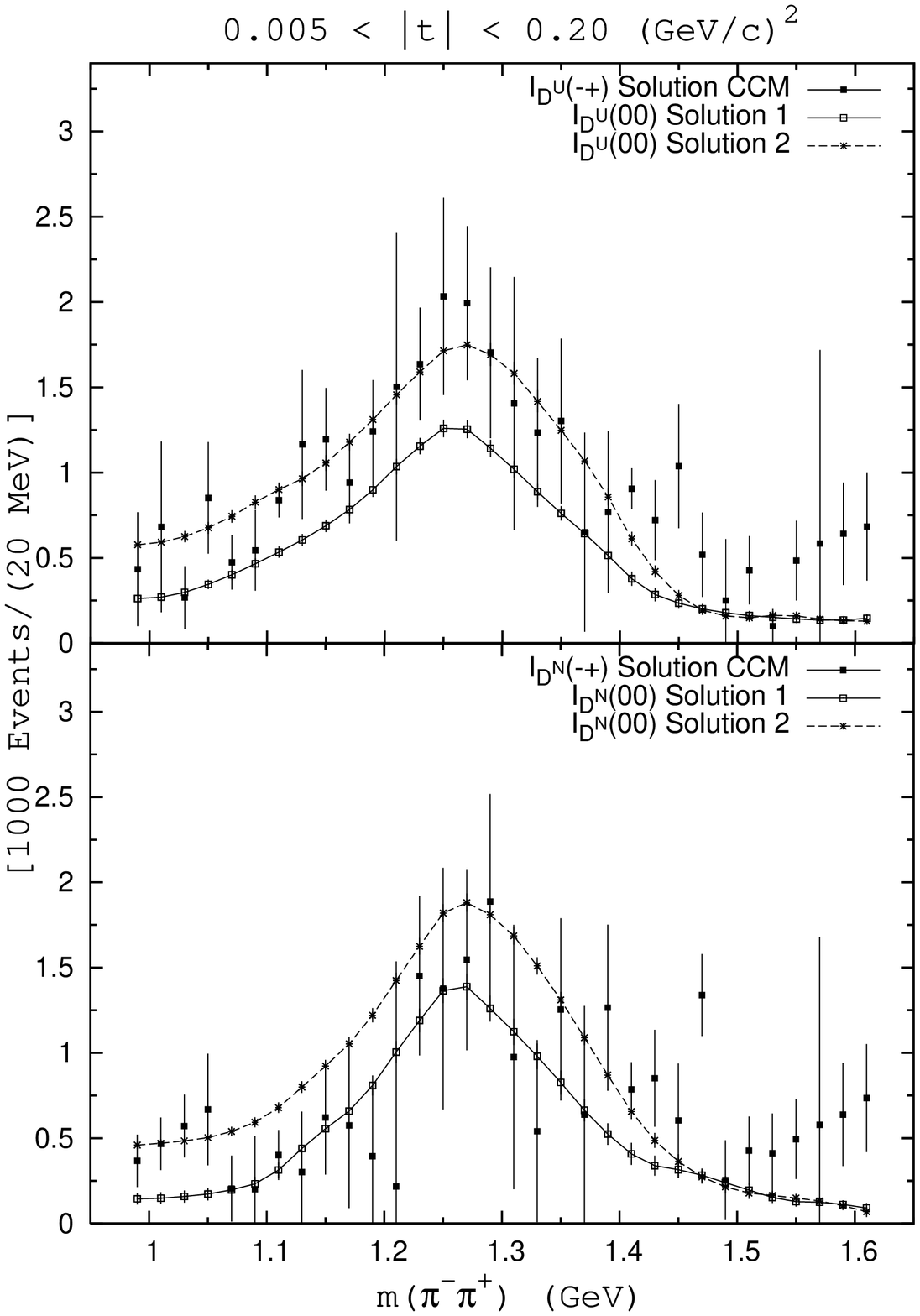}
\caption{$D^U$-wave and $D^N$-wave intensities $I_{D^U}(-+)$ and $I_{D^N}(-+)$ for $\pi^- p \to \pi^- \pi^+ n$ from CERN-Cracow-Munich analysis~\cite{kaminski02} compared to two Solutions 1 and 2 for intensities $I_{D^U}(00)$ and $I_{D^N}(00)$ for $\pi^- p \to \pi^0 \pi^0 n$ from Ref.~\cite{gunter01}. The intensities $I_{D^U}(00)$ and $I_{D^N}(00)$ were interpolated to 20 MeV bins and scaled to 17.2 GeV/c.}
\label{fig3}
\end{figure}

\begin{figure}
\includegraphics[width=14cm,height=12cm]{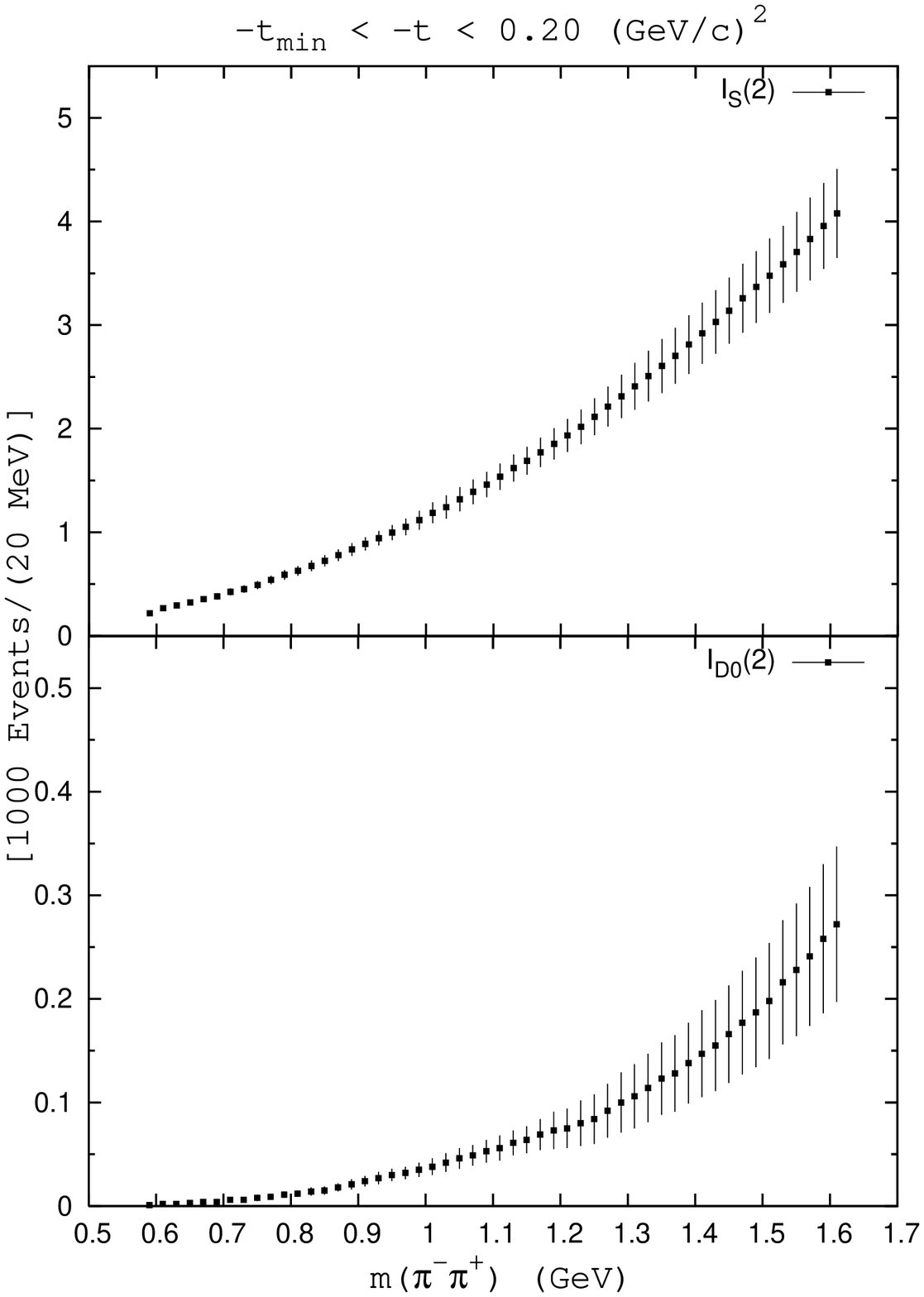}
\caption{Isospin $I=2$ intensities $I_S(2)$ and $I_{D^0}(2)$ from CERN data on $\pi^+ \pi^+$ production at 12.5 GeV/c~\cite{hoogland77} scaled to 17.2 GeV/c.}
\label{fig4}
\end{figure}

For the $\pi^0 \pi^0$ channel we used the BNL data at 18.3 GeV/c~\cite{gunter01}. The BNL data were converted from native BNL units "intensity/40 MeV" into our units "1000 events/20 MeV" using a conversion factor $F=0.6700\times10^{-4}$. We obtained this factor by comparing the $f_2(1270)$ peak value in their Figure 5F given in units "intensity/40 MeV" with the value of coresponding 4 bins at $f_2(1270)$ peak in their Figure 4a given in units "events/10 MeV". The data in two bins $0.01 < |t| < 0.10$ (GeV/c)$^2$ and $0.10 < |t| < 0.20$ (GeV/c)$^2$ were combined by addition to a sigle bin $0.01 < |t| < 0.20$ (GeV/c)$^2$ corresponding to the CERN measurements. The data were then interpolated to 20 MeV bins and scaled to 17.2 GeV/c using phase and flux factor $K(s,m^2)$ given by~\cite{svec97a}
\begin{eqnarray}
K(s,m^2) & = & {G(s,m^2) \over {\text{Flux}(s)}}\\
\nonumber
G(s,m^2) & = & {1 \over {(4 \pi )^3}}{q(m^2) \over {\sqrt{[s-(M+\mu)^2][s-(M-\mu)^2]}}}\\
\nonumber
\text{Flux}(s) & = & 4Mp_{\pi lab}
\end{eqnarray}
where $q(m^2)={1 \over {2}} \sqrt{m^2 -4\mu^2}$ is the pion momentum in the center of mass of the dipion system of mass $m$, and $M$ and $\mu$ are the nucleon and pion mass, respectively. The two Solutions 1 and 2 for intensities $I_A(00)$, $A=S,D^0,D^U,D^N$ are shown in Figures 1, 2 and 3 and compared with the corresponding intensities in $\pi^- p \to \pi^- \pi^+ n$.\\

For the $\pi^+ \pi^+$ channel we used CERN data on $\pi^+ p \to \pi^+ \pi^+ n$ at 12.5 GeV/c~\cite{hoogland77}. This data was used to determine the isospin $I$=2 $S$- and $D$-wave amplitudes $f^{I=2}_{J=0}$ and $f^{I=2}_{J=2}$ in $\pi \pi$ scattering using a pion exchange formula for $S$- and $D^0$-wave helicity flip amplitudes $A_1(++)$
\begin{equation}
A_1(++)=F(s,t,m^2) \sqrt{2J+1}  f^{I=2}_J
\end{equation}
with an assumption that the non-flip amplitudes vanish $A_0(++)$=0. The partial wave intensities $I_S(++)$ and $I_{D^0}(++)$ at 17.2 GeV/c were then reconstructed using
\begin{equation}
I_A(++)=|A_1(++)|^2 \text{  and  } I_A(2)={3 \over {2}} I_A(++)
\end{equation}
The common factor $|F(s,t,m^2)|^2$ was taken from the analysis of Kaminski {\sl et al} ~\cite{kaminski02} by identifying the mean values of our $I_S(2)$ with the values of their $S$-wave $I=2$ contribution $I(2)$ at 17.2 GeV/c and $0.005<|t|<0.20$ (GeV/c)$^2$ presented in their Figure 2. The errors on $I_A(2)$ are given by the errors on $f^{I=2}_{J=0,2}$ and are taken from the CERN analysis in Ref.~\cite{hoogland77}. The results for $I_S(2)$ and $I_{D^0}(2)$ were scaled to 17.2 GeV/c and are shown in Figure 4.\\

The intensities $I_{D^U}(++)$ and $I_{D^N}(++)$ are not known but the CERN measurements of $\pi^+ p \to \pi^+ \pi^+ n$ indicate they are small and consistent with zero. From the  Figures 2 and 3 we see that $I_{D^U}(-+)$ and $I_{D^N}(-+)$ are about 10 \% of the intensity $I_{D^0}(-+)$ at the peak value for both $\pi^- \pi^+$ and $\pi^0 \pi^0$ processes. We therefore expect a similar ratio of $I_{D^U}(++)$ and $I_{D^N}(++)$ to $I_{D^0}(++)$ in $\pi^+ \pi^+$ production and take
\begin{equation}
I_{D^U}(2) = I_{D^N}(2) = I_{D^0}(2) /10
\end{equation}
The errors on $I_{D^U}(2)$ and $I_{D^N}(2)$ were set at 20 \% of errors on $I_{D^0}(2)$. The sensitivity of results on these assumptions was tested by varying the ratios of $I_{D^U}(2)$ and $I_{D^N}(2)$ to $I_{D^0}(2)$ from 10 \% to 100 \%.

\subsection{Signatures of breakdown of Generalized Bose-Einstein symmetry.}

Generalized Bose-Einstein symmetry requires that all three cosines $\cos \Omega_k(A)$, $k=$1,2,3 for all amplitudes $A=S$, $D^0$, $D^U$ and $D^N$ have physical values for all dipion masses for any momentum transfer $t$ within the whole error volume of the measured intensities. A breakdown of the symmetry occurs when $\cos \Omega_k(A)$ calculated from constraints (3.12) and their $D$-waves analogues have unphysical values for some mass range $m_A$ for amplitude $A$ in a large portion of the measured error volume. When that happens, the amplitudes violating Generalized Bose-Eistein symmetry no longer vanish. It is important to note that physical values of the cosines are only consistent with the symmetry but do not prove it. Only the unphysical values of $\cos \Omega_k(A)$ signal conclusively the violation of the symmetry. However, due to possible uncertainties in the data used in the tests some of our results should be viewed with caution.

\subsection{Test of constraints from Generalized Bose-Einstein symmetry.}

The constraints (3.12) arising from the assumption of Generalized Bose-Einstein symmetry were tested as follows. The error volume of the intensities $I_A(-+)$, $I_A(00)$ and $I_A(2)$ was sampled by Monte Carlo method in each mass bin. For each of the 10,000 samplings used in the analysis equations (3.12) were used to calculate the cosines $\cos \Omega_1(A)$, $\cos \Omega_2(A)$ and $\cos \Omega_3(A)$ for all solution combinations for each of the amplitudes $A=S,D^0,D^U,D^N$. A distribution of values of $\cos \Omega_k(A)$ has been obtained for each $k$ and $A$ which defined the range and average value of $\cos \Omega_k(A)$ in each mass bin. In general, these average values of $\cos \Omega_k(A)^{av}$ were close to the mean values of $\cos \Omega_k(A)^*$ calculated from the mean values of the intensities.\\

In each mass bin a number count was taken of the physical and unphysical values of $\cos \Omega_k(A)$ to quantify any possible violation of the constraints (3.12). The program also verified that the non-linear condition (3.13) on the cosines is satisfied for both the physical and unphysical values of $\cos \Omega_k(A)$, $k$=1,2,3 for each Monte Carlo sampling. Importantly, the number counts for unphysical value of $\cos \Omega_k(A)$ were identical for all three cosines in each mass bin. In the Figures 5 - 10 below we thus present only the results for $\cos \Omega_1(A)$.\\

\begin{figure} [ht]
\includegraphics[width=12cm,height=10cm]{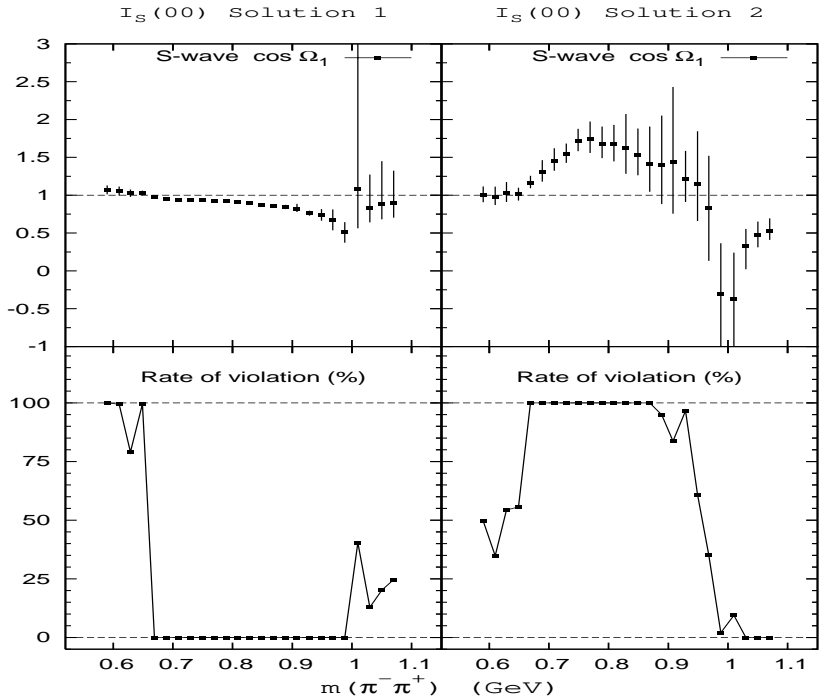}
\caption{Violations of constraints (3.12) in $S$-wave pion pair production below 1080 MeV with Solution (1,1) for $S$-wave intensity $I_S(-+)$ from analysis~\cite{svec07b}. Results from CERN-Cracow-Munich analysis~\cite{kaminski02} are similar.}
\label{fig5}
\end{figure}

\begin{figure} [ht]
\includegraphics[width=12cm,height=10cm]{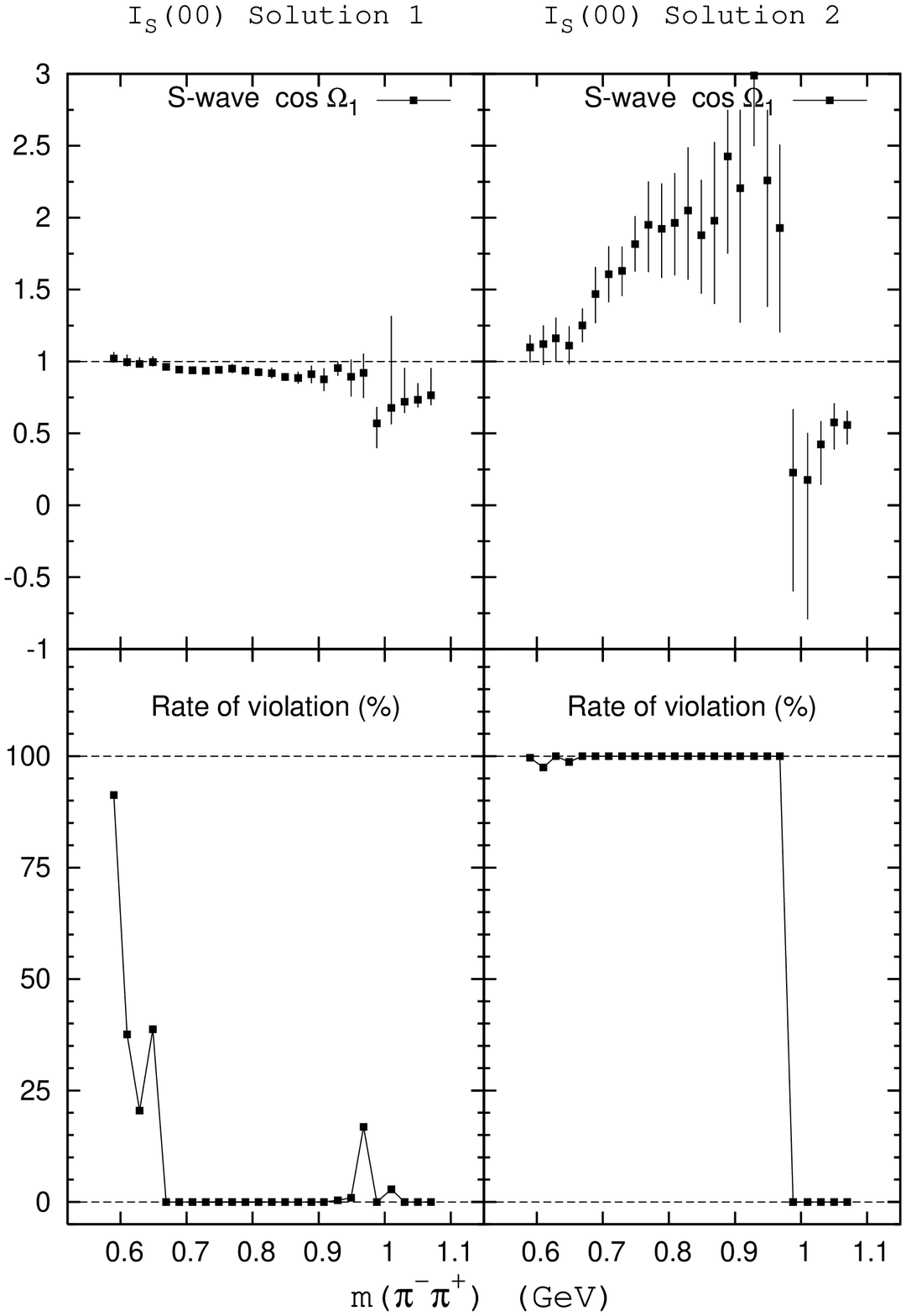}
\caption{Violations of constraints (3.12) in $S$-wave pion pair production below 1080 MeV with Solution (2,2) for $S$-wave intensity $I_S(-+)$ from analysis~\cite{svec07b}. Results from CERN-Cracow-Munich analysis~\cite{kaminski02} are similar.}
\label{fig6}
\end{figure}

\begin{figure}
\includegraphics[width=12cm,height=10cm]{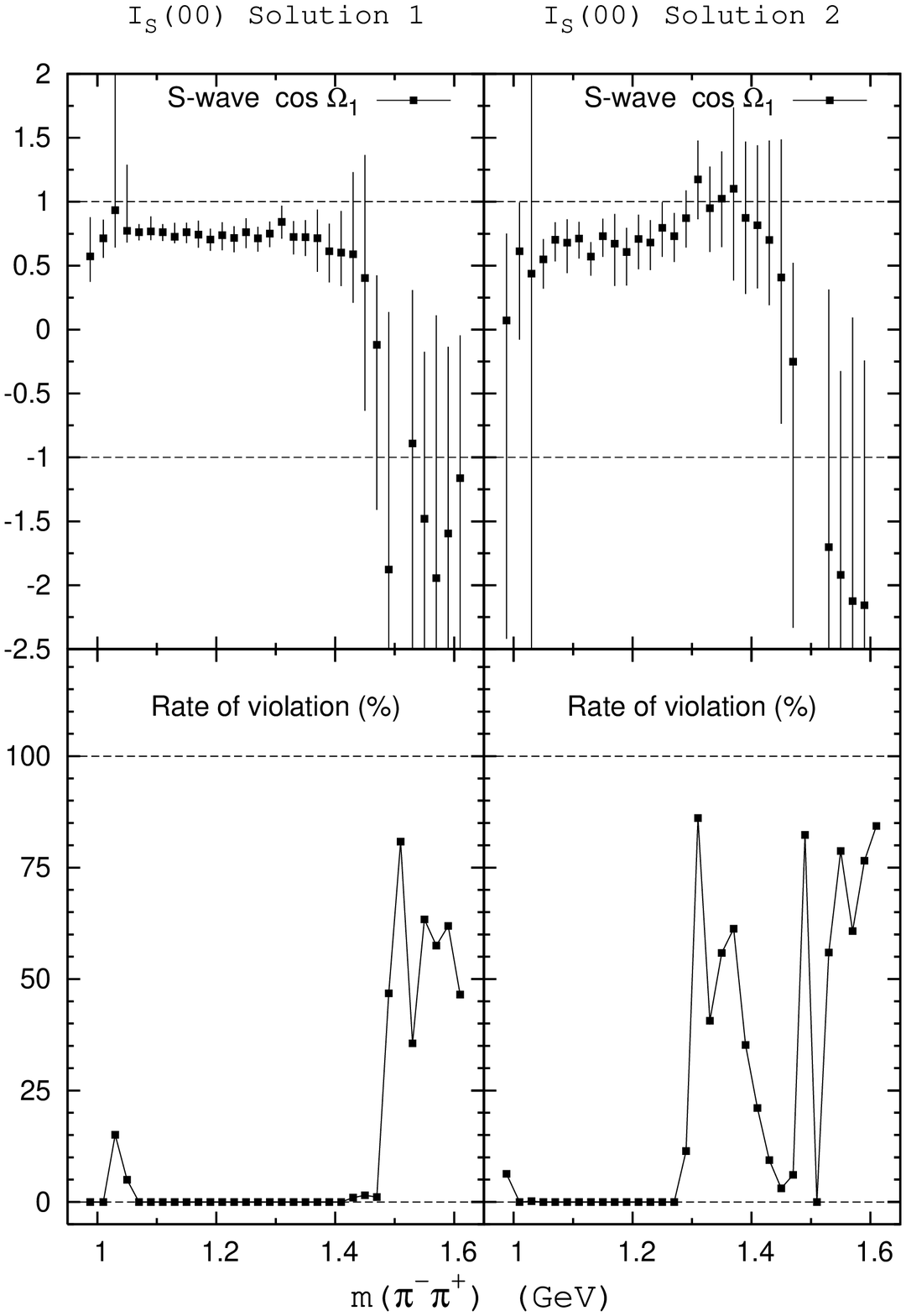}
\caption{Violations of constraints (3.12) in $S$-wave pion pair production in the mass range 980 - 1610 MeV with $S$-wave intensity $I_S(-+)$ from CERN-Cracow-Munich analysis~\cite{kaminski02}.}
\label{fig7}
\end{figure}

\begin{figure}
\includegraphics[width=12cm,height=10cm]{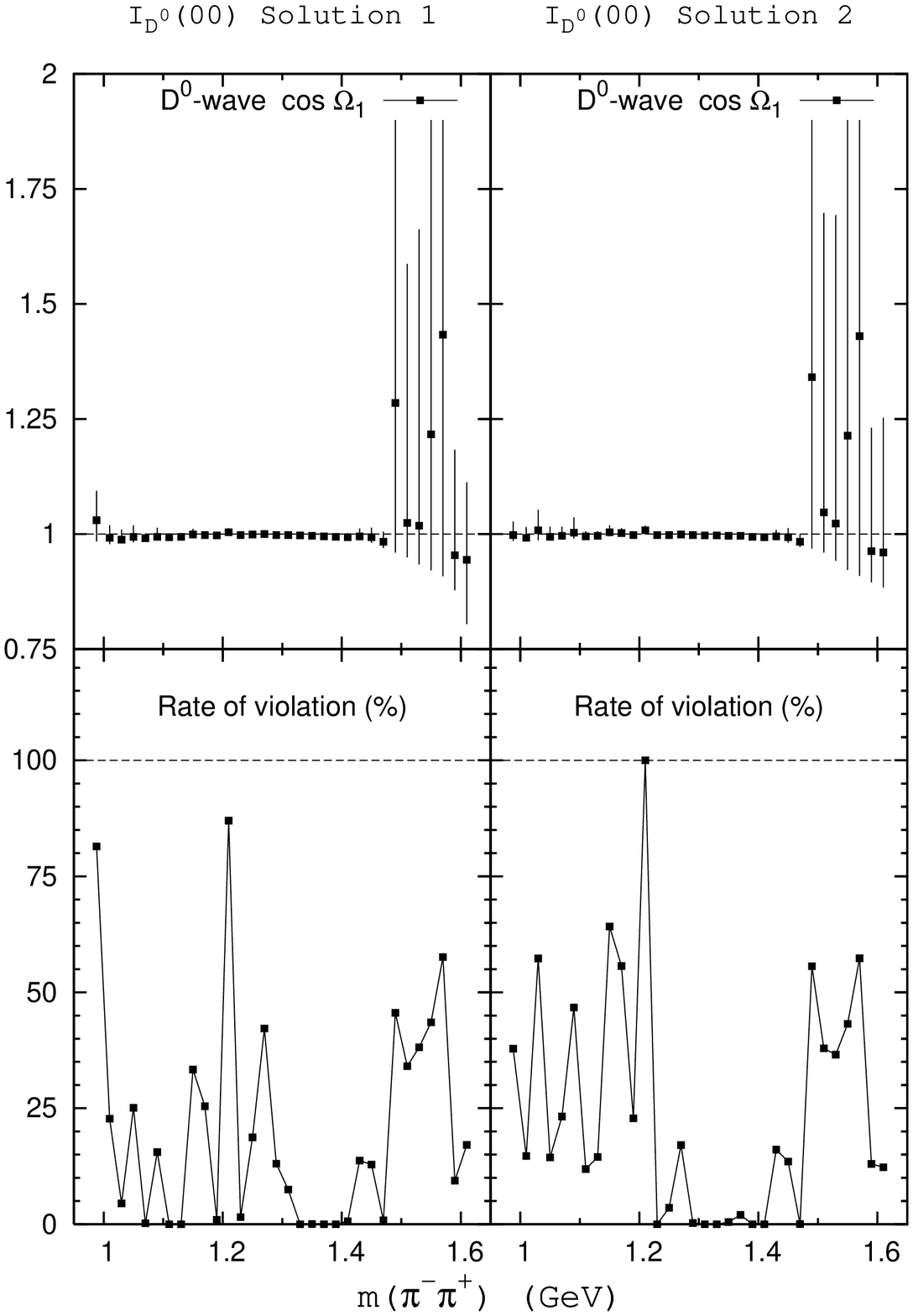}
\caption{Violations of constraints (3.12) in $D^0$-wave pion pair production in the mass range 980 - 1610 MeV with $D^0$-wave intensity $I_{D^0}$ from CERN-Cracow-Munich analysis~\cite{kaminski02}.}
\label{fig8}
\end{figure}

\begin{figure}
\includegraphics[width=12cm,height=10cm]{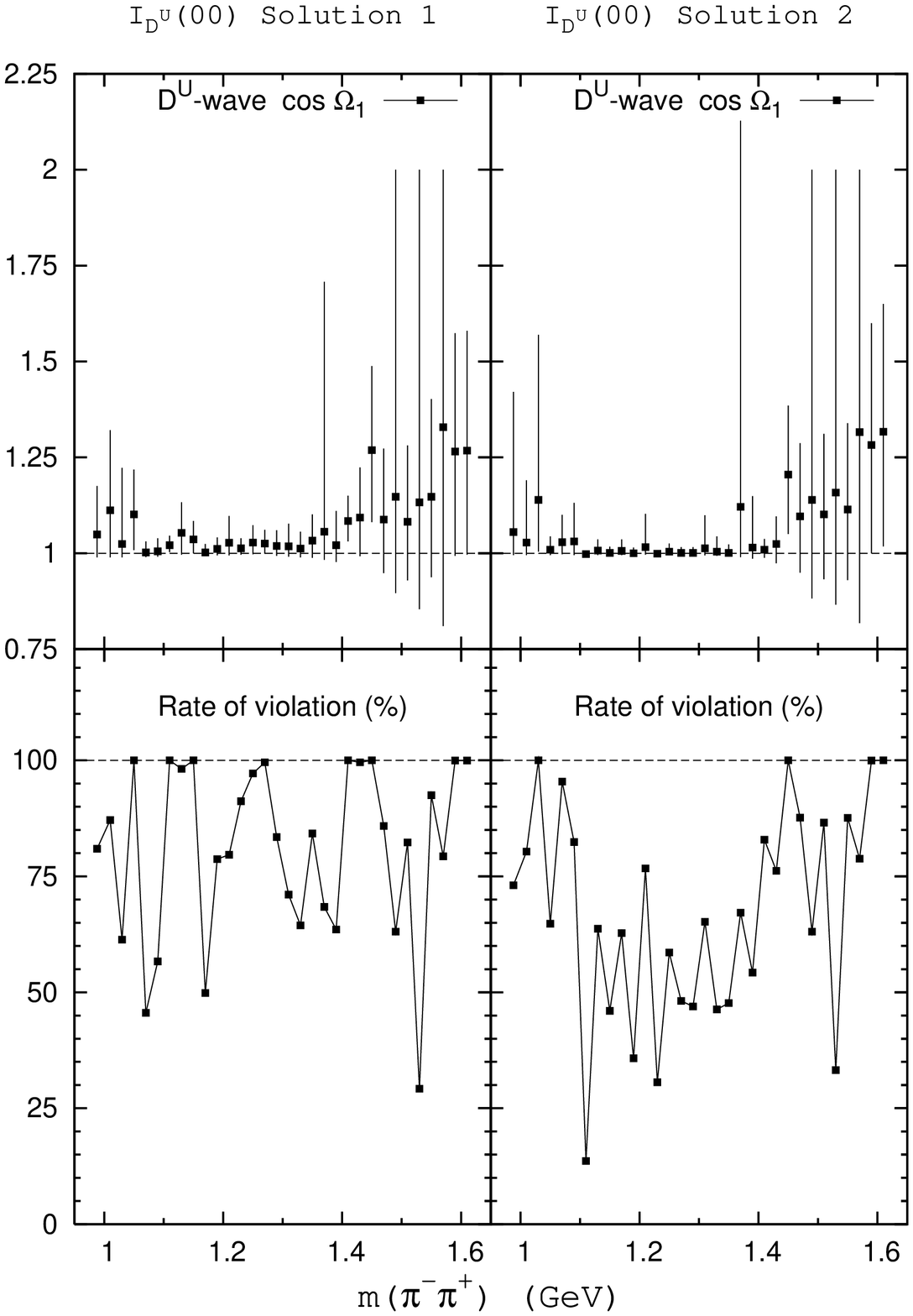}
\caption{Violations of constraints (3.12) in $D^U$-wave pion pair production in the mass range 980 - 1610 MeV with $D^U$-wave intensity $I_{D^U}$ from CERN-Cracow-Munich analysis~\cite{kaminski02}.}
\label{fig9}
\end{figure}

\begin{figure}
\includegraphics[width=12cm,height=10cm]{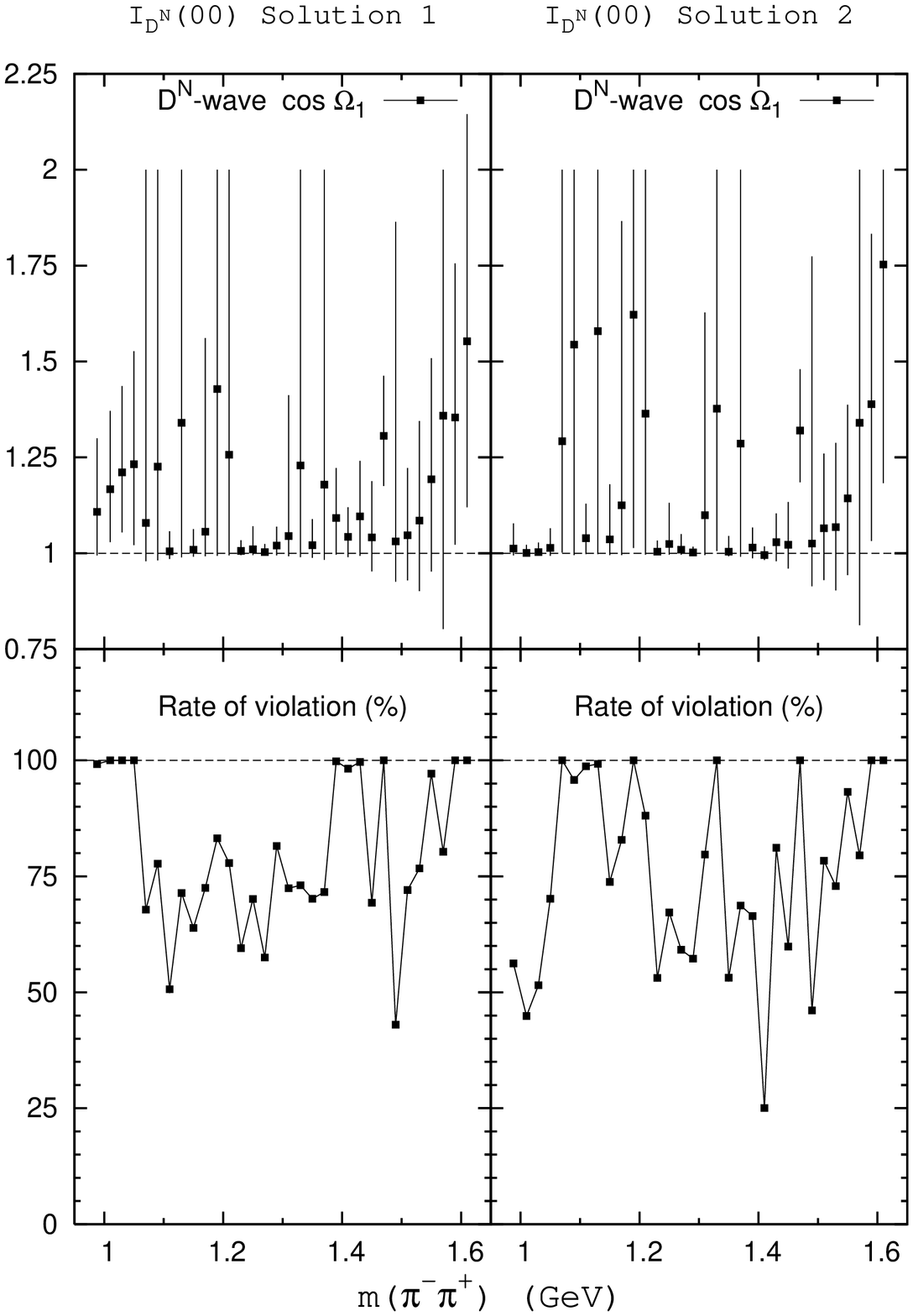}
\caption{Violations of constraints (3.12) in $D^N$-wave pion pair production in the mass range 980 - 1610 MeV with $D^N$-wave intensity $I_{D^N}$ from CERN-Cracow-Munich analysis~\cite{kaminski02}.}
\label{fig10}
\end{figure}

The results for $S$-wave below 1080 MeV are shown in Figures 5 and 6. The results for $S$-, $D^0$-, $D^U$- and $D^N$-waves in the mass range 980-1620 MeV are shown in Figures 7, 8, 9 and 10, respectively. Each Figure shows $\cos \Omega_1(A)$ and the corresponding  fraction of unphysical values of $\cos \Omega_1(A)$ to evaluate the degree of violation of constraints (3.12) or, equivalently, the degree of breaking of Generalized Bose-Einstein symmetry.\\

The results for the two solutions of $S$-wave intensity $I_S(-+)$ below 1080 MeV from our high resolution analysis~\cite{svec07b} and the CCM analysis~\cite{kaminski02} are very similar. Figures 5 and 6 show the results for the Solutions (1,1) and (2,2) of $I_S(-+)$ from our analysis~\cite{svec07b}. For Solution 1 of $I_S(00)$ and for both solutions of $I_S(-+)$ there is a clear violation of the constraints (3.12) for 580 - 680 MeV. For Solution 2 of $I_S(00)$ there is a massive violation of constraints (3.12) in the mass range 600-980 MeV for both solutions of $I_S(-+)$.\\

\begin{figure}
\includegraphics[width=12cm,height=10cm]{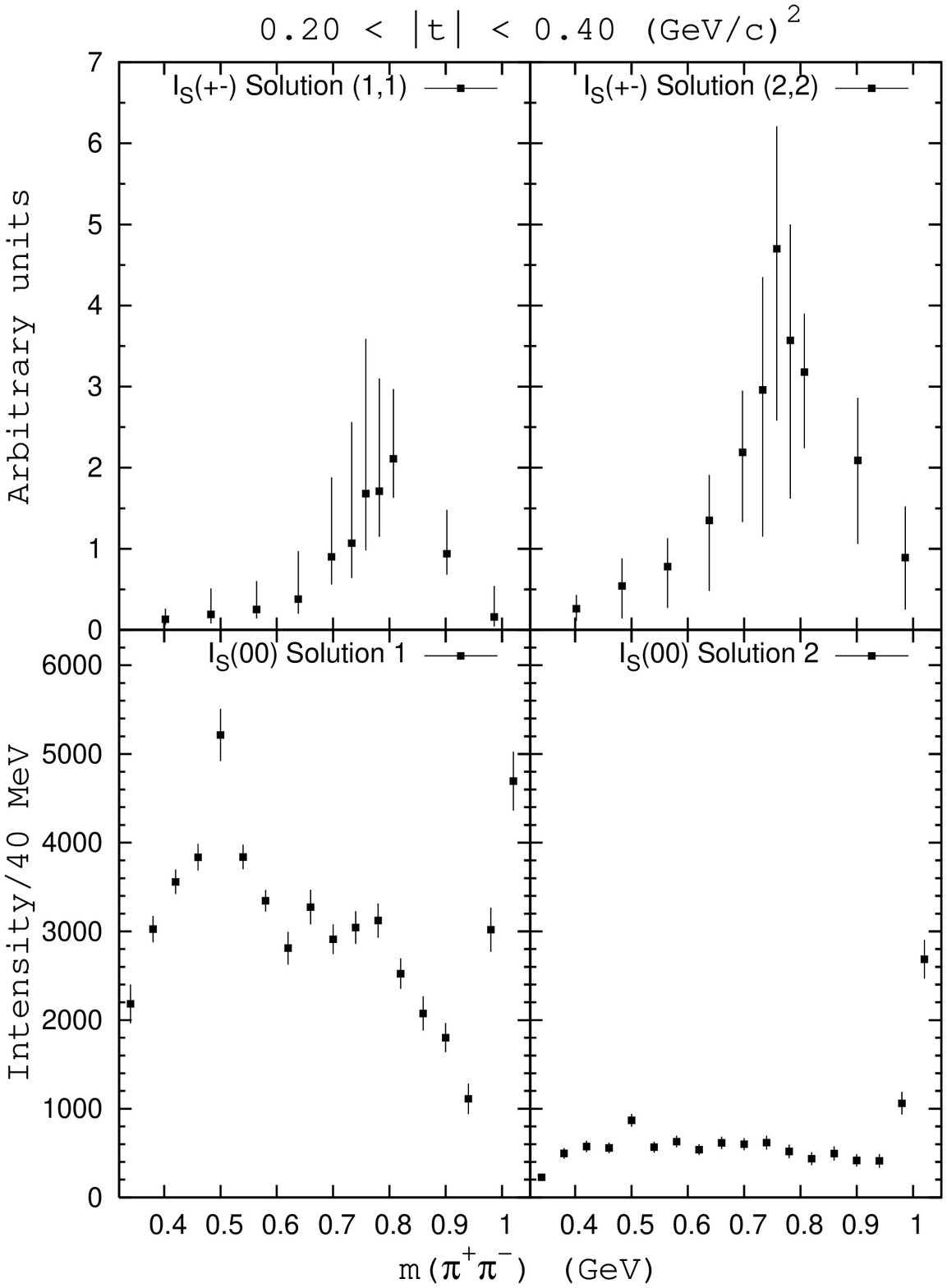}
\caption{Solutions (1,1) and (2,2) for $S$-wave intensity $I_S(+-)$ in $\pi^+ n \to \pi^+ \pi^- p$ at 11.85 GeV/c compared with Solutions 1 and 2 for $S$-wave intensity $I_S(00)$ in $\pi^- p \to \pi^0 \pi^0 n$ at 18.3 GeV/c for momentum transfers $0.20<|t|<0.40$ (GeV/c)$^2$.}
\label{fig11}
\end{figure}

Figure 7 shows the results for $S$-wave in the mass range 980-1620 MeV. The constraints (3.12) are clearly violated at larger masses 1400-1620 MeV in both solutions for $I_S(00)$. Figure 8 presents the results for $D^0$-wave. Interestingly, $\cos \Omega_1(D^0)$ is near 1 but clear violations of constraints (3.12) are again observed in the mass range 1400-1620 MeV for both solutiond of $I_{D^0}(00)$.\\

Figures 9 and 10 present the results for $D^U$- and $D^N$-waves, respectively, assuming $I_{D^U}(2)=I_{D^N}(2)$ is 10 \% of $I_{D^0}(2)$. There is a large violation of consraints (3.12) in all mass bins for both solutions of $I_{D^U}(00)$ and $I_{D^N}(00)$. The sensitivity of this result to the assumptions about $I_{D^U}(2)$ and $I_{D^N}(2)$ was tested by varying their ratio to $I_{D^0}(2)$ from 10 \% to 100 \%. The large violations of constraints (3.12) persist all the way to the 100 \% ratio although the fractions of unphysical values of $\cos \Omega_1(D^U)$ and $\cos \Omega_1(D^N)$ get smaller at higher ratios and, at 100 \%, become zero in some of the mass bins.\\

In Figure 11 we show Solutions (1,1) and (2,2) for $I_S(+-)$ from high resolution amplitude analysis of CERN measurements of $\pi^+ n \to \pi^+ \pi^- p$ on polarized target at 11.85 GeV/c for dipion masses in the range 360 - 1040 MeV at larger momentum transfers $0.20<|t|<0.40$  (GeV/c)$^2$~\cite{svec07b} and compare them with Solutions 1 and 2 for $I_S(00)$ in the same $t$ bin. The intensities $I_S(+-)$ decrease rapidly with decreasing mass below 600 MeV while the Solution 1 for the intensity $I_S(00)$ is large. We may assume that the intensities $I_S(-+)$ for momentum transfers $0.005<|t|<0.20$ (GeV/c)$^2$ will also rapidly decrease below 580 MeV well below the Solution 1 for the intensity $I_S(00)$. From this analogy we may conclude that the violation of the constraints (3.12) will remain and in fact increase below 580 MeV for Solution 1 of $I_S(00)$.

\subsection{Interpretation of the tests.}

Despite some caveats regarding the accuracy of the data used in the tests, the tests reveal a rather convincing case for violation of Generalized Bose-Einstein symmetry and for the presence of amplitudes violating this symmetry in $\pi^- p \to \pi^- \pi^+ n$ reaction. The violation of the generalized Bose-Einstein symmetry is not related to strong interactions or to electromagnetic interactions. It arises from a new $CPT$ violating interaction of the pion creation process with a quantum environment. The presence of $\rho^0(770)$ in the $S$-wave due to this interaction is associated with violations of the constraints by the Solution 1 for $I_S(00)$ below 680 MeV and by Solution 2 for $I_S(00)$ for masses below 980 MeV.\\ 

The mass range of 1400 - 1620 MeV is within the mass region of $\rho^0(1450)$ resonance. The violations of constraints (3.12) in this mass range by $S$- and $D^0$-wave amplitudes suggests the presence of $\rho^0(1450)$ in the $S$- and $D^0$-waves akin to $\rho^0(770)-f_0(980)$ mixing. This result is expected from our model of interaction of the pion creation process with the environment~\cite{svec07b} in which the interaction with the environment induces transitions of $\rho^0(1450)$ into the $S$- and $D^9$-waves. There may be a similar effect in the $D^U$- and $D^N$- waves.

\section{Evidence for quantum entanglement of $\pi^-\pi^+$ isospin states.}

When Generalized Bose-Einstein symmetry is violated the two-pion state $\pi^-\pi^+$ produced in $\pi^- p \to \pi^- \pi^+ n$ needs no longer be a separable state (2.3). In general, it will be an entangled state 
\begin{equation}
|E(\pi^-\pi^+)>=a_S|S>+a_A|A>=a|\pi^-\pi^+>+b|\pi^+\pi^->
\end{equation}
where $|S>$ and $|A>$ are symmetric and antisymmetric $\pi^-\pi^+$ isospin states (2.4), respectively, and
\begin{eqnarray}
a & = {1\over{\sqrt{2}}}(a_S+a_A)\\
\nonumber
b & = {1\over{\sqrt{2}}}(a_S-a_A)
\end{eqnarray}
Only for $a_S=a_A={1\over{\sqrt{2}}}$ are the two-pion states $\pi^-\pi^+$ separable.
The transversity amplitudes then have a general form
\begin{eqnarray}
\text{J=even:} \quad A^{J \eta}_{\lambda,\tau} & = & a_SC^{J \eta}_{\lambda,\tau}+a_AV^{J \eta}_{\lambda, \tau}\\
\nonumber
\text{J=odd:} \quad A^{J \eta}_{\lambda,\tau} & = & a_AC^{J \eta}_{\lambda,\tau}+a_SV^{J \eta}_{\lambda, \tau}
\end{eqnarray}
where $C^{J \eta}_{\lambda,\tau}$ and $V^{J \eta}_{\lambda, \tau}$ are Generalized Bose-Einstein symmetry conserving and violating amplitudes. The entanglement amplitudes $a_S$ and $a_A$ do not depend on transversity $\tau=u,d$ nor on solution labels $i,j=1,2$. We can see that from the general expression for helicity amplitudes in terms of transversity amplitudes~\cite{svec07a,svec07c}
\begin{equation}
A^{J\eta}_{\lambda,n}(ij)={(-i)^n \over{\sqrt{2}}}(A^{J \eta}_{\lambda,u}(i)+(-1)^n A^{J \eta}_{\lambda,d}(j))
\end{equation}
where $n=0,1$ for helicity non-flip and flip amplitudes, respectively. The helicity amplitudes $A^{J,\eta}_{\lambda,n}(ij)$ must share the same isospin state (5.1) as the transversity amplitudes in order for them to be expressed in terms of symmetry conserving and violating helicity amplitudes $C^{J \eta}_{\lambda, n}$ and $V^{J \eta}_{\lambda,n}$ formed by combinations (5.4) of symmetry conserving and violating transversity amplitudes
$C^{J \eta}_{\lambda,\tau}$ and $V^{J \eta}_{\lambda, \tau}$, respectively.\\

To examine the data below 1080 MeV for evidence of quantum entanglement of $\pi^- \pi^+$ isospin states we shall focus on $S$- and $P$-wave amplitudes with helicity $\lambda=0$ which we write in the form
\begin{eqnarray}
S_\tau(i) & = & a_SS^C_\tau(i)+a_AS^V_\tau(i)\\
\nonumber
L_\tau(i) & = & a_AL^C_\tau(i)+a_SL^V_\tau(i)
\end{eqnarray}
where $i=1,2$. The symmetry conserving amplitude $S^C_\tau(i)$ must satisfy relation (3.5)
\begin{equation}
S^C_\tau(i)=S_\tau(00,i)-\sqrt{{3 \over{2}}} S_\tau(++,i)
\end{equation}
where $S_\tau(00,i)$ and $S_\tau(++,i)$, i=1,2 are transversity $S$-wave amplitudes in $\pi^- p \to \pi^0 \pi^0 n$ and $\pi^+ p \to  \pi^+ \pi^+ n$ reactions, respectively. 
From our model for non-unitary dynamics of $\rho^0(770)-f_0(980)$ mixing~\cite{svec07b} we have for the Generalized Bose-Einstein symmetry violationg amplitudes
\begin{eqnarray}
S^V_\tau(i) & \sim & L^C_\tau(i)\\
\nonumber
L^V_\tau(i) & \sim & S^{00}_\tau(i)
\end{eqnarray}
where $S^{00}_\tau(i)$ is the isospin $I=0$ component of $S^C_\tau(i)$. We are interested in the difference
\begin{equation}
S_\tau(1)-S_\tau(2)=a_S(S^C_\tau(1)-S^C_\tau(2))+a_A (S^V_\tau(1)-S^V_\tau(2))
\end{equation}
To proceed we shall make several plausible assumptions. First, we expect that there are very small differences between the $I=2$ amplitudes (which themselves are small below $\sim$ 800 MeV) and set
\begin{equation}
S_\tau(++,1)-S_\tau(++,2)=0
\end{equation}
Then
\begin{equation}
S^C_\tau(1)-S^C_\tau(2)=S_\tau(00,1)-S_\tau(00,2)
\end{equation}
From the data in Figure 1 we expect this difference to be large below $\sim$ 900 MeV. In our next step we note that the amplitudes $|L_\tau(1)|^2$ and $|L_\tau(2)|^2$ are very similar for both transversities $\tau=u,d$~\cite{svec07b}. We thus expect there will be only small differences between the amplitudes $S^V_\tau(1)$ and $S^V_\tau(2)$ and set
\begin{equation}
S^V_\tau(1)-S^V_\tau(2)=0
\end{equation}
From (5.8) we thus have
\begin{equation}
|S_\tau(1)-S_\tau(2)|^2=|a_S|^2|S_\tau(00,1)-S_\tau(00,2)|^2
\end{equation}
In our previous work~\cite{svec07c} we have shown that for physical helicity amplitudes the phases of the two solutions $S_\tau(1)$ and $S_\tau(2)$ must be equal. Then
\begin{equation}
|S_\tau(1)-S_\tau(2)|^2=(|S_\tau(1)|-|S_\tau(2)|)^2
\end{equation}
\begin{figure}
\includegraphics[width=12cm,height=10cm]{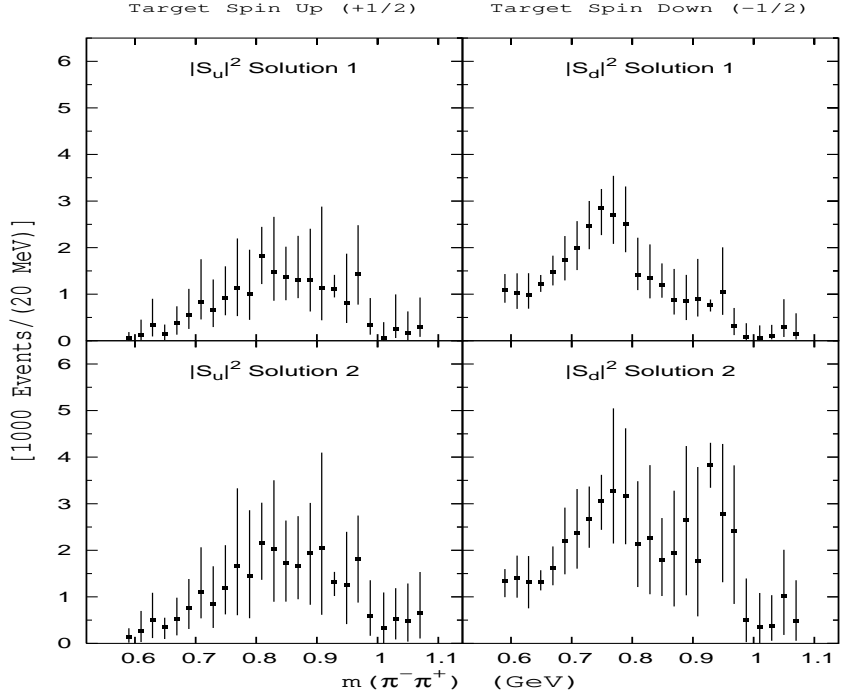}
\caption{Solutions 1 and 2 for $S$-wave amplitudes $|S_u|^2$ and $|S_d|^2$ in $\pi^- p \to \pi^- \pi^+ n$ at 17.2 GeV/c from analysis~\cite{svec07b}.}
\label{fig12}
\end{figure}
\begin{figure}
\includegraphics[width=12cm,height=10cm]{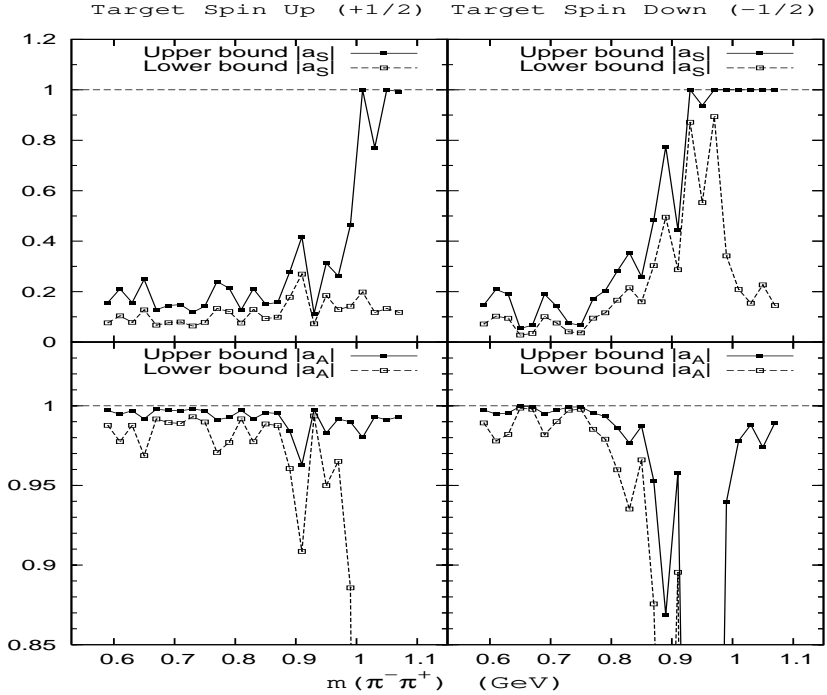}
\caption{Estimates of lower and upper bounds on entanglement amplitudes $|a_S|$ and $|a_A|$ from data on $\pi^- p \to \pi^- \pi^+ n$ and $\pi^- p \to \pi^0 \pi^0 n$ at 17.2 GeV/c.}
\label{fig13}
\end{figure}
This term is known and the moduli squared of the amplitudes are shown in Figure 12 taken from~\cite{svec07b}. Since we do not know the relative phase of the amplitudes $S_\tau(00,1)$ and $S_\tau(00,2)$ we rewrite (5.12) as inequalities
\begin{eqnarray}
|a_S|^2(|S_\tau(00,1)|-|S_\tau(00,2)|)^2 & \leq & (|S_\tau(1)|-|S_\tau(2)|)^2 \\
\nonumber
(|S_\tau(1)|-|S_\tau(2)|)^2 & \leq & |a_S|^2(|S_\tau(00,1)|+|S_\tau(00,2)|)^2
\end{eqnarray}
In our final step we assume that
\begin{eqnarray}
|S_u(00,1)|^2=|S_d(00,1)|^2=I_S(00,1)/2\\
\nonumber
|S_u(00,2)|^2=|S_d(00,2)|^2=I_S(00,2)/2
\end{eqnarray}
where $I_S(00,1)$ and $I_S(00,2)$ are the Solutions 1 and 2 for the $S$-wave intensity in $\pi^- p \to \pi^0 \pi^0 n$ shown in Figure 1. This assumption is in fact an assumption made in BNL data analysis~\cite{gunter01}. From (5.14) we finally obtain lower and upper bounds on the entanglement amplitude $|a_S|$
\begin{equation}
\sqrt{2}{{||S_\tau(1)|-|S_\tau(2)||} \over{\sqrt{I_S(00,1)}+\sqrt{I_S(00,2)}}} \leq |a_S| \leq \sqrt{2}{{||S_\tau(1)|-|S_\tau(2)||} \over{\sqrt{I_S(00,1)}-\sqrt{I_S(00,2)}}}
\end{equation}
From the normalization condition $|a_S|^2+|a_A|^2=1$ we obtain lower and upper bounds on the entanglement amplitude $|a_A|$.\\

The results for entanglement amplitudes $|a_S|$ and $|a_A|$ calculated for both cases of transversity $\tau=u,d$ are shown in Figure 13. To estimate $|a_S|$ we used mean values of $|S_\tau(i)|$ and $I_S(00,i)$, $i=1,2$ and errors were not calculated. Below 840 MeV the results from both transversity calculations are consisten with $|a_S| \sim 0.10-0.20$ and $|a_A| \sim  0.98-0.99$ indicating a quantum entanglement of the $\pi^- \pi^+$ isospin states in $\pi^- p \to \pi^- \pi^+ n$. \\

The effect of the quantum entanglement below 840 MeV is a suppression of the conserving amplitude $S^C_\tau(i)$ and the clear dominance of the violating amplitude $S^V_\tau(i)$ in the mass range of $\rho^0(770)$ resonance in the $S$-wave amplitudes $S_\tau(i)$. For dipion masses 980-1080 MeV calculations with both transversities are again consistent and suggest larger values of $|a_S|$. This value of $|a_S|$ leads to large contribution from the violating amplitude $L^V_\tau(i)$ to the amplitudes $L_\tau(i)$ which manifests itself by the presence of a dip in $|L_\tau(i)|^2$ at $\sim $980 MeV corresponding to $f_0(980)$ resonance. There is a divergence between the two calculations in the mass range 840-980 MeV which suggests that some of our simplifying assumptions are violated. This violation may be also responsible for the large range of values of the entanglement amplitudes above 980 MeV.\\

We cannot estimate the entanglement amplitudes above 980 MeV using the CERN-Cracow-Munich analysis~\cite{kaminski02}. The $S$- and $P$-wave subsystem of the reduced density matrix measured on transversely polarized target is always analytically solvable and yields two solutions for each transversity amplitude $A_u(i),A_d(j), i,j=1,2$~\cite{svec07b}. While CERN-Cracow-Munich analysis used the analytical solutions below 980 MeV, above this mass  fits were made to measured moments which are linear combinations of density matrix elements. These fits produced a single solution for the amplitudes. This can happen if the analytical solutions are close and cannot be distinguished by the fits. The single solution for the $S$-wave amplitudes obtained in the fits cannot be used in (5.16) to estimate the entanglement amplitudes.

\section{Conclusions.}

We have derived three linearly independent constraints on partial wave intensities with even spin in $\pi^- p \to \pi^- \pi^+ n$, $\pi^- p \to \pi^0 \pi^0 n$ and $\pi^+ p \to \pi^+ \pi^+ n$ reactions that must be satisfied by the data if the Generalized Bose-Einstein symmetry holds true. Available data violate the constraints. The violation of the symmetry implies the presence of the symmetry violating contributions to transversity amplitudes in $\pi^- p \to \pi^- \pi^+ n$ and allows for quantum entanglement of the $\pi^- \pi^+$ isospin states. We have derived approximate expressions for lower and upper bounds on entanglement amplitudes. The bounds provide clear experimental evidence for quantum entanglement of $\pi^- \pi^+$ isospin states below 840 MeV and suggest the entanglement at higher dipion masses. The small values of $|a_S| \sim 0.10-0.20$ below 840 MeV explain the puzzling difference between the $S$-wave intensities in $\pi^- p \to \pi^- \pi^+ n$ and $\pi^- p \to \pi^0 \pi^0 n$ and reveal a suppression of isospin $I=0,2$ contribution in the $S$-wave amplitudes in $\pi^- p \to \pi^- \pi^+ n$. The large isospin $I=1$ contribution of $\rho^0(770)$ to both $S$- and $P$-wave amplitudes is due to large entanglement amplitude $|a_A| \sim 0.98-0.99$.\\

A crucial aspect of the our tests was the treatment of all solutions for the intensities as valid physical solutions. For instance, when looking at the tests of the constraints (3.12) for the $S$-wave amplitudes below 1080 MeV it is not enough to look only at the Solution 1 for $I_S(00)$ or to look at the two solutions for $I_S(00)$ separately. Instead, the tests for the two solutions must be viewed and interpreted together since the underlying amplitudes for different solutions form a single system.\\

The violations of Generalized Bose-Einstein symmetry and the entanglement of $\pi^- \pi^+$ isospin states are not related to strong or electromagnetic interactions. We suggest that they provide new evidence for the existence of a quantum environment and its $CPT$ violating interaction with pion creation processes. The origin of the environment remains an open question.\\

The new interaction manifests itself in several ways. It leads to splitting of the transversity amplitudes $A_u$ and $A_d$ to two physical solutions $A_u(i), i=1,2$ and $A_d(j), j=1,2$ which results in mixed final state (1.4) and in a $CPT$ violating non-unitary evolution of pure initial states to mixed final states in all $\pi N \to \pi \pi N$ processes~\cite{svec07a,wald80}. The splitting reflects the interacting degrees, or the quantum numbers, of the environment described by quantum states $|i>|j>$.\\

The interaction also leads to isospin conserving transitions between resonant $q \overline{q}$ modes with spin $K$, helicity $\mu$ and isospin $I_K$ produced in $\pi^- p \to q \overline{q} n$ and two-pion states with spin $J$, helicity $\lambda$ and isospin $I(\pi \pi)=I_K$. As the result of these transitions resonant modes can "leak" into different partial wave amplitudes~\cite{svec07b,svec07c}. The transitions are observable as violations of Generalized Bose-Einstein symmetry, $\rho^0(770)-f_0(980)$ mixing~\cite{svec07b} and the presence of $\rho^0(1450)$ in $S$- and $D^0$-wave amplitudes. Transitions from $K=2$ $q \overline{q}$ mode to $J=0$ $\pi^0 \pi^0$ state could also explain the large differences in the two solutions for the $S$-wave amplitudes in $\pi^- p \to \pi^0 \pi^0 n$ below 900 MeV.\\

The $CPT$ violating interaction with the environment produces entangled two-pion isospin state $|E(\pi^- \pi^+)>=a_S|S>+a_A|A>$. A separable state $\pi^- \pi^+$ is expected from strong interactions and Generalized Bose-Einstein symmetry. The entanglement of the $\pi^- \pi^+$ state is another aspect of $CPT$ violation in $\pi^- p \to \pi^- \pi^+n$. The entangled state is not experimentally preparable since the entanglement amplitudes are not constants but may depend on energy $s$, dipion mass $m$ and momentum transfer $t$. As the result the amplitude for the $CPT$ conjugate process is not defined and $CPT$ symmetry looses its meaning in the interaction of pion creation process with the quantum environment.\\

Our study has been limitted to partial wave intensities since the reactions $\pi^- p \to \pi^0 \pi^0 n$ and $\pi^+ p \to \pi^+ \pi^+ n$ were measured only on unpolarized targets. Future experiments may measure all pion creation processes on polarized targets and thus enable more detailed studies of quantum entanglement and non-unitary dynamics of pion creation processes on the level of amplitudes, deepening our understanding of this new class of phenomena.

\acknowledgements
I wish to acknowledge the use of CERN-Cracow-Munich analysis of $\pi^- p \to \pi^- \pi^+ n$ and $I(2)$ intensity presented in~\cite{kaminski02} and of BNL data on $\pi^- p \to \pi^0 \pi^0 n$ located at World Wide Web~\cite{pi0pi0pwa}.

\end{document}